\documentclass[12pt]{article}
\usepackage{amssymb}
\usepackage{mathrsfs}
\usepackage{bm}
\usepackage{graphicx}
\usepackage{subfigure}
\usepackage{hyperref}
\begin{document}
\begin{titlepage}
\begin{center}
\begin{Large}
{\bf The hidden fluxes, that control the fluctuations of scalar fields}
\end{Large}

\vskip0.5truecm

Stam Nicolis\footnote{E-Mail: Stam.Nicolis@lmpt.univ-tours.fr, stam.nicolis@idpoisson.fr}

\vskip0.5truecm

{\sl Institut Denis Poisson, Université de Tours, Université d'Orléans, CNRS (UMR7013)\\
Parc Grandmont, 37200, Tours, France}

\end{center}
\begin{abstract}
The fluctuations of scalar fields, that are invariant under rotations of the worldvolume, in Euclidian signature, can be described by a system of Langevin equations. These equations can be understood as defining a change of variables in the functional integral for the noise, with which the physical degrees of freedom are in equilibrium. The absolute value of the Jacobian of this change of variables therefore repackages the fluctuations. This provides a new way of relating the number and properties of scalar fields with the consistent and complete description of their fluctuations and is another way of understanding the relevance of supersymmetry, which, in this way, determines the minimal number of real scalar fields (e.g. two in two dimensions, four in three dimensions and eight in four dimensions), in order for the system to be consistently closed. 

The classical action of the scalar fields, obtained in this way, contains a surface term and a remainder, in addition to the canonical kinetic and potential terms. The surface term describes possible flux contributions in the presence of boundaries, while the remainder describes additional interactions, that can't be absorbed in a redefinition of the canonical terms. It is, however, through its  combination with the surface term that the noise fields can be recovered, in all cases. However their identities can be subject to anomalies. 

What is of particular, practical, interest  is the identification of the noise fields, as functions of the scalars, whose correlation functions are Gaussian. This implies new identities, between the scalars, that can be probed in real, or computer, experiments.
\end{abstract}
\end{titlepage}
\section{Introduction}\label{intro}
In ref.~\cite{parisi_sourlas} Parisi and Sourlas studied the ${\mathcal N}=2,D=2$ Wess--Zumino model in the following way. Their starting point was the  generalization of the Langevin equation
\begin{equation}
\label{genlangevin}
\eta_I(x) = \sigma^\mu_{IJ}\frac{\partial\phi_J}{\partial x_a}+\frac{\partial W}{\partial\phi_I}
\end{equation}
where the $\sigma^\mu$ generate a Clifford algebra
\begin{equation}
\label{clifford}
\left\{\sigma^\mu,\sigma^\nu\right\}=2\delta^{\mu\nu}
\end{equation}
(in Euclidian signature). They can, thus,  be identified with the Pauli matrices, in $D=2$ and a convenient choice is $\sigma^1=\sigma_x$ and $\sigma^2=\sigma_z.$ What is important is that the $\sigma^\mu$ realize a Majorana representation, i.e. have only real entries, since both sides  of eq.(\ref{genlangevin}) are real numbers.

The $\eta_I(x)$ are Gaussian fields, with ultra--local 2--point function, i.e. $\langle\eta_I(x)\rangle=0,\langle\eta_I(x)\eta_J(x')\rangle=\delta_{IJ}\delta(x-x')$ with the multipoint correlation functions given by Wick's theorem. These properties can be expressed in equivalent form by the partition function
\begin{equation}
\label{Zeta}
Z=\int\,[\mathscr{D}\bm{\eta}(x)]\,e^{-\int\,d^2\bm{x}\,\frac{1}{2}\eta_I(x)\eta_J(x)\delta^{IJ}}\equiv 1
\end{equation}
by a suitable choice of units. Of course any noise distribution can be used, provided its partition function exists and it can be reconstructed from its correlation functions. The reason it's useful to use a Gaussian distribution is to avoid higher derivative terms and the reason it's useful to focus on Gaussian distributions with ultra--local 2--point function is for the resulting action of the scalars to be local (in applications to magnets these requirements can be modified). 

Now the observation of Parisi and Sourlas is that eq.~(\ref{genlangevin}) can be understood as the injunction to perform the indicated change of variables in eq.~(\ref{Zeta}) and obtain the partition function for the scalar fields $\phi_I(x):$
\begin{equation}
\label{Zphi}
Z = \int\,[\mathscr{D}\phi_I]\,\left|\mathrm{det}\,\frac{\partial\eta_I}{\partial\phi_J}\right|\,e^{-\int\,d^2\bm{x}\,\frac{1}{2}\left( \sigma^\mu_{IJ}\frac{\partial\phi_J}{\partial x_\mu}+\frac{\partial W}{\partial\phi_I}   \right)\left(  \sigma^\nu_{KL}\frac{\partial\phi_L}{\partial x_\nu}+\frac{\partial W}{\partial\phi_K}\right)\delta^{IK}}=1
\end{equation}
The partition function remains equal to 1, since we have, simply, changed variables in the partition function for the noise. The question of potential anomalies, due to regularization, can be addressed by  the study of the identities that the noise fields, expressed in terms of the scalars, satisfy. 

The absolute value of the Jacobian thus captures all of the fluctuations of the classical action. An interesting question, that was, indeed, one of the tenets of ref.~\cite{parisi_sourlas}, is, whether the fluctuations of the classical action, in turn, can produce the absolute value of the determinant, that can, therefore, be identified as the Jacobian, that realizes the change of variables from the scalar fields to the noise fields. If this is the case, this would mean that the canonical partition function of the scalars, in fact, is the Witten index. 

This statement wasn't expressed in these terms in ref.~\cite{parisi_sourlas}, though it's clearly a logical consequence of the paper, but in ref.~\cite{Nicolis:2014yka,Nicolis:2016osp} for the case of the non-relativistic particle and in ref.~\cite{Nicolis:2017lqk} for the $D=2,\mathcal{N}=2$ Wess--Zumino model, on the basis of numerical simulations, where the identities that the noise fields, $\eta_I(x),$ expressed in terms of the scalars, by eq.~(\ref{genlangevin}), are expected to satisfy.  What these results, also, imply is that it isn't necessary to deal with the full action in eq.~(\ref{Zphi}), that contains fermions--and the phase factor, $e^{\mathrm{i}\theta_\mathrm{det}},$ that's a non--local function of the scalars--but it suffices to compute the correlation functions of $\eta_I(x),$ expressed in terms of the scalars, by eq.~(\ref{genlangevin}), sampled using the action of the scalars only.  This map, between the scalars and the noise fields, is known as the Nicolai map~\cite{Nicolai:1980jc,Nicolai:1980js}; however Nicolai introduced it for supersymmetric theories. In fact it expresses the property that the physical system, described by the RHS, is consistently ``closed'', since its fluctuations can produce the absolute value of the Jacobian to variables, for which the partition function is equal to 1.

Stated more explicitly: The fundamental insight of~\cite{parisi_sourlas}--mentioned, however,  in passing in that paper--is that there is an equivalence between the property that the partition function, defined by eq.~(\ref{Zphi}) is, in fact, the Witten index and the property that the RHS of the Langevin equation~(\ref{genlangevin}) (i.e.   the Nicolai map), defines a Gaussian field, with ultra--local 2--point function, when sampled using the action of the scalars; this is the meaning of the statement that  the superpartners, along with the appropriate phase  of the determinant, are generated by the fluctuations in such a way as to define a closed system.  This represents, of course, a considerable simplification for performing calculations on the lattice,  since it's much easier to perform numerical simulations of scalar theories than of theories that explicitly include fermions. So it is necessary to check that this does occur. First tests were, indeed, consistent with this hypothesis for one and two--dimensional worldvolumes~\cite{Nicolis:2014yka,Nicolis:2017lqk}. 

However, when writing the ``bosonic'' part of the classical action of the Wess--Zumino model, with extended supersymmetry, it is interesting to note that it isn't identical to the corresponding expression obtained upon realizing the change of variables defined by the Nicolai map; in the latter case, additional terms appear. These terms could be shown to be surface terms in $D=2,$ but not for the more interesting cases of $D=3$ and $D=4.$ 
This has been interpreted~\cite{parisi_sourlas,Cecotti:1982ad,Cecotti:1983up} as an obstruction to understanding supersymmetry as  describing the fluctuations of the bosonic part. 

In this note we wish to argue that the obstruction arises because of the assumption that the ``usual terms'' exhausted the possibilities of realizing supersymmetry. What the Nicolai map really implies is that the additional terms do not, in fact, provide an obstruction, but new ways for probing the relevance of supersymmetry in theories, where scalar fields can describe interesting phenomena. In particular, these indicate that one scalar field is a semi--classical approximation and target space Lorentz invariance (which in Euclidian signature is rotation invariance) implies the existence of additional scalar fields; and even more, when a Majorana representation isn't possible (i.e. when $D\equiv\hskip-0.4truecm/\,\,2\,\mathrm{mod}\,8$). Indeed, the exact number is the same as in supersymmetric extensions of the Standard Model, just how it is obtained is different from the usual approach.

This is illustrated in the case of the $\mathcal{N}=2,D=2$ Wess--Zumino model, upon expanding out the classical action for the scalars:
\begin{equation}
\label{Sclass}
S[\phi_I]=\int\,d^2\bm{x}\,\left\{\partial_\mu\phi_I\partial_\nu\phi_J\delta_{\mu\nu}\delta_{IJ}+\frac{1}{2}\frac{\partial W}{\partial\phi_I}\frac{\partial W}{\partial\phi_J}\delta_{IJ}+
\sigma_{IJ}^\mu\partial_\mu\phi_J\frac{\partial W}{\partial\phi_I}
\right\}
\end{equation}
We recognize the canonical kinetic term and the canonical scalar potential for the scalars, in the form expected from supersymmetry; 
but we, also, find an additional term, the ``cross term''. This term may be a total derivative in the continuum (and an infinite series of total derivatives and irrelevant terms (proportional to positive powers of the lattice spacing)--on the lattice).  

But it may, also, apparently, contribute to the equations of motion directly.

To this end, it's useful to write it in the following way:
\begin{equation}
\label{extraterm}
\sigma_{IJ}^\mu\partial_\mu\phi_J\frac{\partial W}{\partial\phi_I}=\partial_\mu\left\{\sigma_{IJ}^\mu\phi_J\frac{\partial W}{\partial\phi_I}\right\}-
\phi_J\sigma_{IJ}^\mu\partial_\mu\frac{\partial W}{\partial\phi_I}=
\partial_\mu\left\{\sigma_{IJ}^\mu\phi_J\frac{\partial W}{\partial\phi_I}\right\}-
\phi_J\partial_\mu\phi_K\sigma_{IJ}^\mu\frac{\partial^2 W}{\partial\phi_I\partial\phi_K}
\end{equation}
The first term is a total derivative in the continuum and, thus, doesn't contribute to the equations of motion, upon imposing periodic boundary conditions. It can contribute, in the presence of boundaries, of course. (This is the flux term, mentioned in the title.) 

Let us focus on the second term. It has the form that suggests it can be written as 
$$
\mathrm{Tr}\left[J\cdot A\right]
$$
where
\begin{equation}
\label{JandA}
\begin{array}{l}
\displaystyle
[J_\mu]_{JK}=\phi_J\partial_\mu\phi_K\\
\displaystyle
\left[A_\mu\right]_{JK}=\sigma_{IJ}^\mu\frac{\partial^2 W}{\partial\phi_I\partial\phi_K}
\end{array}
\end{equation}
Since it isn't a total derivative, by construction, if it doesn't vanish, it will contribute to the equations of motion. This is what was noticed in refs.~\cite{Cecotti:1982ad,Cecotti:1983up}. 
While $[J_\mu]_{JK}$ does have the form of the current that can be constructed from $D$ scalar fields, whether $[A_\mu]_{JK}$ can, indeed, be identified as a gauge field in its own right is, of course, more questionable, since just what the corresponding ``gauge transformations'' might correspond to, isn't clear.  

We would like to find the conditions that make it vanish in various dimensions. If this is possible, we recover a conventional supersymmetric theory,  once we have introduced the Jacobian in the action, using anticommuting fields, 
\begin{equation}
\label{Zsusy}
\begin{array}{l}
Z=\int\,[\mathscr{D}\phi_I]\,[\mathscr{D}\psi_I]\,[\mathscr{D}\chi_I]\,e^{\mathrm{i}\theta_\mathrm{det}}\,e^{-\int\,d^2\bm{x}\,\left\{
\partial_\mu\phi_I\partial_\nu\phi_J\delta_{\mu\nu}\delta_{IJ}+\frac{1}{2}\frac{\partial W}{\partial\phi_I}\frac{\partial W}{\partial\phi_J}\delta_{IJ}
-\psi_I\left(\sigma_{IJ}^\mu\partial_\mu+\frac{\partial^2 W}{\partial\phi_I\partial\phi_J}\right)\chi_J\right\} }=1
\end{array}
\end{equation}
However another possibility is that it doesn't vanish, but becomes a total derivative itself. The reason can be understood in the simplest case, $D=1:$
\begin{equation}
\label{extratermD=1}
\begin{array}{l}
\dot{\phi}W'(\phi)=\frac{dW}{d\tau}=\frac{d}{d\tau}\left(\phi W'\right)-\phi\frac{d}{d\tau}W'=\frac{d}{d\tau}\left(\phi W'\right)-\phi\dot{\phi}W''=\\
\frac{d}{d\tau}\left(\phi W'\right)-\frac{d}{d\tau}\left(\frac{\phi^2}{2}\right)W''
\end{array}
\end{equation}
Here we notice that the remainder (the second term of the last line) doesn't vanish, while the cross term is, in fact, a total derivative; but we shall show that it can become a total derivative:
$$
\begin{array}{l}
\frac{d}{d\tau}\left(\frac{\phi^2}{2}\right)W''=\frac{d}{d\tau}\left(\frac{\phi^2}{2}W''\right)-\frac{\phi^2}{2}\dot{\phi}W'''=\frac{d}{d\tau}\left(\frac{\phi^2}{2}W''\right)-\frac{d}{d\tau}\left(\frac{\phi^3}{3!}\right)W'''=\\
\hskip2.6truecm
\frac{d}{d\tau}\left( \frac{\phi^2}{2}W''-\frac{\phi^3}{3!}W'''\right)+\frac{\phi^3\dot{\phi}}{3!}W^{(4)}
\end{array}
$$
By induction we may readily prove that 
\begin{equation}
\label{extratermD=1induction}
\frac{d}{d\tau}\left(\frac{\phi^2}{2}\right)W''=\frac{d}{d\tau}\left\{   
\sum_{n=2}^\infty\,(-)^n\frac{\phi^n}{n!}W^{(n)}(\phi)
\right\}
\end{equation}
For a polynomial superpotential the series, of course, terminates. 

We remark that a property that played a crucial role in making the remainder a total derivative was that 
$$
\phi^n\dot{\phi}=\frac{d}{d\tau}\left(\frac{\phi^{n+1}}{(n+1)}\right)
$$
This property doesn't hold for more than one fields. So in $D=2$ another way of ``eliminating'' the cross term is used, namely by imposing that the superpotential be a holomorphic function of the fields. While this property is, indeed, very useful for performing calculations analytically, it's not that crucial for numerical work. It does provide motivation for trying to understand how holomorphicity of the superpotential can emerge in the ``classical'' limit. 

In the next section we shall, therefore, recall what happens to the cross  term in $D=2$; then we shall discuss another way of understanding why  there appears to be an obstruction in generalizing the construction for $D>2$ and how the obstruction can be evaded.  The payoff will be new identities between correlation functions, that can be probed in real and computer experiments. 

\section{The cross term in $D=2$}\label{remD2}
The reason the presence of the ``cross term'' is a source of concern is that, while the action for the noise is invariant under global SO(2) transformations, $\delta\eta_I = \theta\varepsilon_{IJ}\eta_J,$ (that correspond to Lorentz boosts upon Wick rotations), it doesn't seem obvious that the Nicolai map(s)~(\ref{genlangevin}) respect this symmetry--i.e. that they imply that $\delta\phi_I=f(\theta)\varepsilon_{IJ}\phi_J.$ While the canonical kinetic and potential terms are, manifestly, invariant under such transformations, the cross term requires more work. 

Indeed, we  notice that the SO(2) transformations act on the indices $I, J$ of the fields--and that $\sigma_{IJ}^\mu$ carries such indices, also. Therefore the cross term is invariant, provided the SO(2) transformations mix non--trivially with transformations of the Pauli matrices and the worldvolume coordinates. This mixing is, of course, allowed and is, indeed, required. This lays to rest any concerns that Lorentz invariance might be explicitly broken and, thus, doesn't single out holomorphic superpotentials as the only allowed by rotation invariance. But it does indicate that more care is needed in monitoring the identities the correlation functions of the scalars should satisfy. The results of the numerical simulations, that extend the work reported in ref.~\cite{Nicolis:2017lqk}, will be reported in future publications. 

Therefore the cross term doesn't signal the explicit breaking of supersymmetry, but its possible spontaneous breaking, that can be probed by computing the 1--point functions of the noise fields, 
\begin{equation}
\label{1ptnoise}
\left\langle\eta_I\right\rangle=\left\langle\sigma_{IJ}\frac{\partial\phi_J}{\partial x^\mu}+\frac{\partial W}{\partial\phi_I}\right\rangle\stackrel{?}{=}0
\end{equation}
The breaking can be anomalous if the 2--point functions of the noise fields, expressed in terms of the scalars, aren't $\delta-$functions of the worldvolume arguments:
\begin{equation}
\label{2ptnoise}
\left\langle\left(\eta_I(\bm{x})-\left\langle\eta_I(\bm{x})\right\rangle\right)\left(\eta_J(\bm{x}')-\left\langle\eta_J(\bm{x}')\right\rangle\right)\right\rangle\stackrel{?}{=}\mathrm{const}\times\delta_{IJ}\delta(\bm{x}-\bm{x}')
\end{equation}
The simulations in ref.~\cite{Nicolis:2017lqk} show that, while the error bars are quite large, it is possible to get reasonable results that suggest that supersymmetry isn't broken, although lattice artifacts can be seen.

Of course the higher order connected correlation functions must vanish to numerical precision; and the  challenge here is storing the datasets that are needed for the analysis. 
 
\section{The obstruction for $D>2$ and how it can be evaded}\label{obsev}
If we try to study the case of a worldvolume of more than two dimensions in the same way, we come across a problem, that makes focusing on the remainder term, in the context presented previously, simply, wrong.

In the literature the problem was presented in different guises: 
 
Parisi and Sourlas, in ref.~\cite{parisi_sourlas} tried to generalize the approach to $D>2,$ as did Cecotti and Girardello in ref.~\cite{Cecotti:1982ad,Cecotti:1983up}. The argument  that defined an obstruction, for Parisi and Sourlas, was that, in $D>2,$ it didn't seem possible to define holomorphicity of the superpotential--that was the property that implied that the cross term is a surface term--in a unique way.

For Cecotti and Girardello,the observation was that, for $D>2$ the remainder  does contribute to the equations of motion, in the continuum; in addition, its presence in the lattice was shown to be inconsistent with the symmetries of the lattice action, especially when fermions were explicitly taken into account. 

However the statement about an obstruction does deserve further analysis. One reason is that the property that the remainder doesn't  vanish, in any event, doesn't preclude $\eta_I(x)$ from being Gaussian fields, with ultra--local 2--point function, once the boundary term is taken into account. 

But this remark misses the more severe problem, namely,  that the real reason for the difficulty in $D>2,$ is the absence of a Majorana representation for the Clifford algebra, when $D\neq 2\,\mathrm{mod}\,8.$ This is the real problem, whose solution requires a doubling of the degrees of freedom--and, therefore, precludes a consistent closure only using the ``original'' degrees of freedom. 

So let us discuss why doubling is necessary and what it entails.

It's necessary, since the RHS of the equations~(\ref{genlangevin}) isn't real, if the $\sigma_{IJ}^\mu$ don't all have real entries, even if they're Hermitian. Therefore the complex conjugate quantities are, also, required, viz.
\begin{equation}
\label{genlangevindagger}
\eta_I^\dagger(\bm{x})=\sigma_{JI}^\mu\frac{\partial\phi_J^\dagger}{\partial x^\mu}+\left(\frac{\partial W}{\partial\phi_I}\right)^\dagger
\end{equation}
(since $[\sigma_{IJ}^\mu]^\dagger=\sigma_{JI}^\mu$) 
The reason is that, if the $\sigma^\mu$ have imaginary entries, then eq.~(\ref{genlangevin}) doesn't define a useful change of variables in the partition function of the noise, since the $\eta_I$ don't take, only, real values. Also, that, if the RHS of eq.~(\ref{genlangevin}) doesn't take real values, the action isn't positive definite (it's not even real). 

In order for the change of variables to make sense at all, it's necessary to take into account the additional degrees of freedom, by considering the corresponding generalization of the partition function of the noise, viz.
\begin{equation}
\label{noisecompl}
Z=\int\,[\mathscr{D}\eta][\mathscr{D}\eta^\dagger]\,e^{-\int\,d^D\bm{x}\,\eta_I^\dagger(\bm{x})\eta_I(\bm{x})}=1
\end{equation}
and to perform the change of variables, from the $\eta_I,\eta_I^\dagger$ to the $\phi_I,\phi_I^\dagger$ as imposed by eqs.~(\ref{genlangevin}) and~(\ref{genlangevindagger})~\footnote{It is interesting that it is this approach that was used in ref.~\cite{parisi_sourlas} to deduce that holomorphicity of the superpotential implied that the cross term was a surface term. }

In this way, also, the action will be real--valued and bounded from below. 

So we have to deal with Gaussian distributions of complex numbers, rather than real numbers--this means, in particular, that the argument of the complex numbers doesn't decouple from their modulus (this was stressed by Cecotti and Girardello~\cite{Cecotti:1982ad,Cecotti:1983up}). 

In $D=3,$ where all three--hermitian--Pauli matrices are needed, the consistent description requires, therefore two complex scalars, i.e. four real scalars. It will be interesting to 
study the new conformal  theories that can be defined in this way. 

Similarly, in $D=4,$ that's more relevant for particle physics, upon choosing all $\gamma-$matrices to be hermitian, $\{\gamma_\mu,\gamma_\nu\}=2\delta_{\mu\nu}$ in Euclidian signature, the impossibility of defining a Majorana representation implies that not all matrices can have real entries. Therefore it's necessary to double the degrees of freedom 
and the consistent description requires four complex scalars, i.e. eight real scalars. It's interesting to remark that this is the  number of scalars required by the minimal supersymmetric extension of the Standard Model--but found by a quite different approach. 

What this construction implies, is new relations between the correlation functions of the scalars, expressing the property that the $\eta_I$ and $\eta_I^\dagger$ define  Gaussian fields with ultra--local 2--point function. 

This is the most ``visible'' novelty of this approach, for making contact with experiment, real or on the computer. 

\section{Conclusions}\label{concl}
The discovery of the Higgs boson at the LHC has made the consistent  theoretical understanding of the fluctuations of scalar fields of topical interest. While supersymmetry was always understood as describing the degrees of freedom that represent the fluctuations, including them explicitly, beyond perturbation theory, has been a non--trivial task.
The stochastic approach~\cite{parisi_sourlas} not only shows that the terms, that are the most difficult to handle in numerical simulations, don't need to be taken into account directly, but can be probed, indirectly, more efficiently than might seem  obvious, but, also, implies that the Nicolai map isn't relevant only for manifestly supersymmetric theories, but for any theory and, indeed, it is supersymmetry that expresses how a physical system, in equilibrium with its fluctuations, can be described as consistently closed. In this way of thinking, the breaking of supersymmetry implies that the system, in fact, is open. Indeed, the way scalar fields appear in the Standard Model is coupled to gauge fields; and the construction of the Nicolai map for gauge fields is, still, work in progress~\cite{Ananth:2020lup,Ananth:2020gkt,Lechtenfeld:2021yjb}. 

Monitoring the identities of the noise fields can, also, provide insight in how anomalies may appear, that have been the subject of recent investigations~\cite{Katsianis:2019hhg,katsianis2020supersymmetry}. 

Finally, another domain, where scalar fields and their fluctuations appear is in cosmology and it will be of interest to see how the stochastic approach might be of relevance for describing the fluctuations of the inflaton~\footnote{Cf. the talk of K. Skenderis at this conference.}

{\bf Acknowledgements:} I would like to thank the organizers of ``HEP2021--The 38th Conference on Recent Developments in High Energy Physics and Cosmology'' of the Hellenic Society for the Study of High Energy Physics for the opportunity to give a talk and for the organization that led to fruitful exchanges, even though they had to be conducted over a medium  that isn't, usually, the first choice. 
Discussions with M. Axenides, E.  Floratos and J. Iliopoulos are gratefully acknowledged.

\bibliographystyle{utphys}
\bibliography{SUSY}
\end{document}